\documentclass[runningheads]{llncs} 
\usepackage[utf8]{inputenc}
\usepackage[T1]{fontenc}
\usepackage[usenames,dvipsnames,svgnames,x11names,table]{xcolor}
\usepackage{cite}
\usepackage{amsmath,amssymb,amsfonts}
\usepackage{algorithmic}
\usepackage{graphicx}
\usepackage{textcomp}
\usepackage{listings}
\usepackage{url}
\usepackage{pifont}
\usepackage[colorlinks=true,linkcolor=blue,urlcolor=blue,citecolor=blue,bookmarks=false]{hyperref}% always load last
\usepackage[abbreviations]{foreign}
\usepackage{booktabs} %used for \toprule
\def\BibTeX{{\rm B\kern-.05em{\sc i\kern-.025em b}\kern-.08em
    T\kern-.1667em\lower.7ex\hbox{E}\kern-.125emX}}	
\usepackage{multirow}
\usepackage[abbreviations]{foreign}
\usepackage[inline]{enumitem}
\usepackage{cleveref}% load after hyperref
\usepackage{subcaption}
\usepackage[subtle]{savetrees}
\usepackage[margin=0pt,skip=6pt,belowskip=0pt,font={small,stretch=0.9},labelfont=bf]{caption}
\usepackage{tabularx, makecell, booktabs}
\usepackage{tikz}
\usepackage{eso-pic} 

\newboolean{showcomments}
\setboolean{showcomments}{true}
\ifthenelse{\boolean{showcomments}}
{ \newcommand{\mynote}[3]{
   \fbox{\bfseries\sffamily\scriptsize#1}
   {\small$\blacktriangleright$\textsf{\emph{\color{#3}{#2}}}$\blacktriangleleft$}}}
{ \newcommand{\mynote}[3]{}}

\newcommand{\xcite}[1]{} % do not cite in shortened paper

\usepackage[font={small},skip=1pt,belowskip=3pt,labelfont=bf]{caption}
\usepackage{listings}
\usepackage[export]{adjustbox}
\usepackage{eso-pic}
\usepackage[most]{tcolorbox}
\usepackage{hyperxmp}

\definecolor{backcolour}{rgb}{0.95,0.95,0.92}
\lstdefinestyle{yaml} {
    columns=fixed,
    numbers=left,                   % where to put the line-numbers
    stepnumber=1,                   % the step between two line-numbers. If it is 1 each line will be numbered
    numbersep=6pt,                  % how far the line-numbers are from the code
    numberstyle=\footnotesize\ttfamily,      % the size of the fonts that are used for the line-numbers
    xleftmargin=11pt,
	framextopmargin=1pt,
	aboveskip=0pt,
    identifierstyle=\color{black}\ttfamily,
    stringstyle=\color{purple}\ttfamily,
    keywords=[1]{platform, architecture, cpu, os,"passthru-devices",volumes},
    keywordstyle=[1]\color{blue}\ttfamily,
    keywords=[2]{features, hardware},
    keywordstyle=[2]\color{NavyBlue}\ttfamily,
    keywords=[3]{SGX, sse4, GPU},
    keywordstyle=[3]\color{blue!50!purple}\ttfamily
}	

\graphicspath{{./plots/}}

\begin{document}

\title{Understanding Cryptocoins Trends Correlations}
\titlerunning{Understanding Cryptocoins Trends Correlations} 
\author{Pasquale De Rosa \orcidID{0000-0001-9726-7075} \and Valerio Schiavoni \orcidID{0000-0003-1493-6603}}
\institute{
University of Neuch\^atel, Switzerland,
\email{first.last@unine.ch}
}

\maketitle

\begin{abstract}
\vspace{-20pt}	
Crypto-coins (also known as cryptocurrencies) are tradable digital assets.
Notable examples include Bitcoin, Ether and Litecoin.
Ownerships of cryptocoins are registered on distributed ledgers (\ie, blockchains).
Secure encryption techniques guarantee the security of the transactions (transfers of coins across owners), registered into the ledger.
Cryptocoins are exchanged for specific trading prices.
While history has shown the extreme volatility of such trading prices across all different sets of crypto-assets,
it remains unclear what and if there are tight relations between the trading prices of different cryptocoins.
Major coin exchanges (\ie, Coinbase) provide trend correlation indicators to coin owners, suggesting possible acquisitions or sells. However, these correlations remain largely unvalidated.

In this paper, we shed lights on the trend correlations across a large variety of cryptocoins, by investigating their coin-price correlation trends over a period of two years.
Our experimental results suggest strong correlation patterns between main coins (Ethereum, Bitcoin) and alt-coins.
We believe our study can support forecasting techniques for time-series modeling in the context of crypto-coins.
%We exploit such correlations to understand the accuracy of state-of-the-art forecasting techniques for time-series modeling (\ie, ARIMA, LSTM, GRU) for the average price of correlated cryptocoins.
%Our results show up to \vs{FIX} accuracy in forecasting the average price of a given alt-coin with a forecasting prediction window of \vs{FIX}.
We release our dataset and code to reproduce our analysis to the research community.

\end{abstract}
\vspace{-20pt}

%!TEX root = paper.tex

\def\confname{22nd International Conference on Distributed Applications and Interoperable Systems (DAIS'22)}
\def\confyear{2022}
\def\confdoi{XXX}

% Copyright text: https://www.ieee.org/publications/rights/rights-policies.html
% and in the Springer's contributor consent PDF.
\definecolor{yellowPaper}{HTML}{fff8ae}
\AddToShipoutPictureFG*{%
  \AtTextUpperLeft{%
    \adjustbox{raise=59pt,center}{
    \begin{tcolorbox}[width=1.5\textwidth,colback=yellowPaper,enhanced,frame hidden,sharp corners]  
        \centering\scriptsize
        \copyright~\confyear\ Springer. Personal use of this material is permitted.
        Permission from Springer must be obtained for all other uses, in any current or future media, including reprinting/republishing this material for advertising or promotional purposes, creating new collective works, for resale or redistribution to servers or lists, or reuse of any copyrighted component of this work in other works.
        This is the author's version of the work.
        The final authenticated version is available online at \href{https://doi.org/10.1007/978-3-031-16092-9_3}{doi.org/10.1007/978-3-031-16092-9\_3} % will be available online
        and has been published in the proceedings of the 
        \confname.
     \end{tcolorbox}} 
  }%
}%

\hypersetup{
    pdfcopyright={\copyright~\confyear\ Springer. Personal use of this material is permitted. Permission from Springer must be obtained for all other uses, in any current or future media, including reprinting/republishing this material for advertising or promotional purposes, creating new collective works, for resale or redistribution to servers or lists, or reuse of any copyrighted component of this work in other works.
    This is the author's version of the work. The final authenticated version is available online at https://doi.org/10.1007/978-3-031-16092-9_7 % will be available online
    and has been published in the proceedings of the 
    \confname.}
}

%!TEX root = paper.tex
\vspace{-6pt}
\section{Introduction}\label{sec:introduction}
\vspace{-6pt}

Cryptocurrencies, also known as crypto-coins, are tradable digital assets, backed by secure encryption techniques to ensure the security of transactions (typically, the transfer of coins across wallets).
Notable examples include Bitcoin~\cite{nakamoto2008bitcoin}, Ether (the native cryptocurrency of the Ethereum blockchain~\cite{buterin2013ethereum}) or Litecoin (used in a fork of the original Bitcoin network).
Nowadays there exists thousands of cryptocurrencies (CoinMarketCap~\cite{coinmarketcap} lists 10039 coins as of April 2022).
Cryptocoins are designed to be traded as a form of digital money: the first useful Bitcoin transaction was used by a peer-to-peer payment between Satoshi Nakamoto (Bitcoin's founder) and one of its early adopters, and dates back to 2009.\footnote{\url{https://www.blockchain.com/btc/block/170}}
Cryptocoins are nowadays traded over online (centralized or decentralized) \emph{exchanges}, including Coinbase~\cite{coinbase}, Kraken~\cite{kraken}, Binance~\cite{binance}, Uniswap~\cite{uniswap}, \etc.
With a current estimated worldwide market-cap of 1.71 Trillion dollars, the cryptocoins economy roughly match the GDP of South Korean in 2021~\cite{IMF}.

\begin{figure}[!t]
	\begin{center}
		\includegraphics[scale=0.7]{{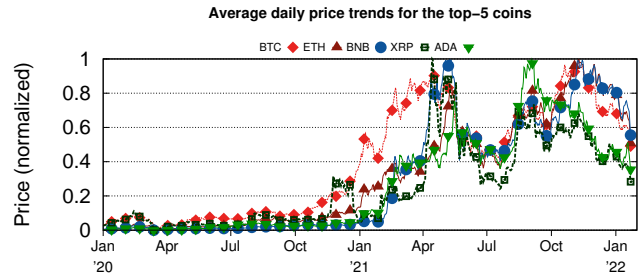}}
	\end{center}
	\vspace{-10pt}
\caption{\label{fig:avg-10}Normalized (min-max) average prices of the top-5 cryptocoins since January 2020.}
	\vspace{-20pt}
\end{figure}

The ownership of cryptocoins is registered on distributed ledgers (\ie, blockchains), together with the corresponding transactions (transfers of coins across wallets).
Cryptocoins are exchanged (\ie, sold, bought) for specific trading prices.
While it is beyond the scope of this work to understand the exact nature of those prices, history has shown the extreme volatility of such trading prices across all different sets of crypto-assets.
For instance, Figure~\ref{fig:avg-10} shows the normalized average daily prices of 5 popular cryptocoins (\ie, BTC, ETH, BNB, XRP, ADA) since January 2020.

It remains unclear what and if there are tight relations between the trading prices of different cryptocoins.
Major cryptocoin exchanges, in particular given the enormous popularity that such digital assets have grown into the large public, and further  facilitated by the easy access to these markets via mobile apps, started to provide \emph{trend correlation} indicators to coin/wallet owners.
Such correlation indicators can possibly drive end-users towards acquisitions or sells. 
%Figure~\vs{show it for coinbase app or another one}.
Coinbase, among the most popular cryptocoin exchanges, indicates the price correlation as \textit{the tendency of other asset prices to change at the same time as the asset shown on the page.} 
In their case, correlation is computed leveraging the Pearson correlation with USD order books over the last 90 days.
%\footnote{\url{https://help.coinbase.com/en/coinbase/getting-started/crypto-education/price-page-information-}} 

However, the nature of such correlations, their intensity as well as the evolution of the correlations through time, remain largely unvalidated.

The \textbf{contributions} of this work-in-progress paper are twofold.
First, we extract the trading prices, as well as other exchange metadata (\eg, open and closing price, market cap, volume), for the top-100 cryptocoins since the last two years from a popular cryptocoin monitoring web-site. 
Second, we leverage this dataset to carry out our preliminary study of the trend correlations between and across crypto-coins. 
Specifically, we investigate daily, weekly and monthly correlation patterns exhibited by two principal cryptocoins, \ie BTC and ETH, against the remaining set of \emph{alt-coins} in our dataset.
Our analysis show strong correlations between the observed trends.
We will leverage these observations in our future work, where we plan to exploit the observed correlations to forecast the future trading trends and by considering the problem of time-series forecasting applied to the crypto-coin market.

We follow an \textit{open science} approach: our datasets will be released and made available to the research and open-source community.

\textbf{Roadmap.} This paper is organized as follows.
Section~\ref{sec:background} provides background materials on Bitcoin, Ethereum, as well as general notions of correlation analysis.
Section~\ref{sec:eval} describes our dataset, as well as our work-in-progress analysis.
%We present our experimental results in \S\ref{sec:eval}.
We briefly cover related work in Section~\ref{sec:relwork}, before concluding and presenting our future work in Section~\ref{sec:conclusion}.

%!TEX root = paper.tex
\vspace{-6pt}
\section{Background}\label{sec:background}
\vspace{-6pt}
\textbf{Cryptocoins in a nutshell.}
%\vs{explain what cryptocoins are in general, what are bitcoin and ether in particular, what are the alt-coins, and other high-level concepts}
Cryptocoins are digitally encrypted assets. 
They were typically designed to replace fiat currencies and used mostly in peer-to-peer networks.
Depending on the incentive natures of the underlying blockchain, cryptocoins (or token) are rewarded to nodes in the network.
We differentiate between three main types of cryptocoins: \emph{(i)} Bitcoin, \emph{(ii)} alt-coins, and \emph{(iii)} stable coins.\footnote{Some characterizations define stable coins as sub-classes of alt-coins, together with secure tokens, utility tokens, and more. We leave as future work to study in-depth the correlations between such sub-types of alt-coins.}
Alt-coins are \emph{alternative} coins to Bitcoin.
Notable examples include Ether (ETH), Cardano (ADA), Litecoin (LTC), or Ripple (XRP). 
A stablecoin is a class of cryptocurrencies that attempt to offer price stability and are backed by a reserve asset, \eg gold or the value of the American dollar. 
Examples include USDT (Tether) and USDC.

\textbf{Time series analysis.}\label{subsec:tsa}
A time series is an $n$-tuple of observations collected sequentially over time. 
Common examples of time series include trends of interest rates and stock prices, daily high and low temperatures, the electrical activity of the heart, \etc.  
The purpose of time series analysis is generally twofold: \emph{(i)} to understand the mechanisms and the inner dynamics of an observed series, and \emph{(ii)} forecast the future values of the series based on the historical ones.
To analyze time series as sequences of random variables (\ie stochastic processes), it is common to assume their stationarity: a time series is  \emph{stationary} if the probability laws that govern its behavior do not change, and its mean $\mu$ is constant over time. 
%The state of the art in stationary time series modeling is represented by the 
The Autoregressive Moving Average is a state-of-the-art stationary time series modeling approach, which combines an Autoregressive (AR) process of order $p$ and of a Moving Average (MA) process of order $q$.
 
Real applications do not expose stationary trends. 
The Autoregressive Integrated Moving Average model (ARIMA) differentiates a nonstationary process a number $d$ of times, until it becomes stationary. 
It is common to observe the presence of a seasonality in the trend of a time series, especially in applications where cyclical tendencies are very common (like business or economics). 
To handle periodical components, a common model is the Seasonal Autoregressive Integrated Moving Average (SARIMA), that can be mapped to a standard ARIMA model in absence of seasonality~\cite{tsanalysis}.

\begin{figure}[!t]
	\begin{center}
		\includegraphics[scale=0.7]{{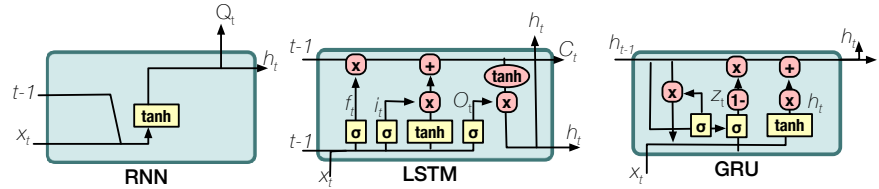}}
	\end{center}
	\caption{\label{fig:rnns}ML-based time series forecasting approaches: RNN, LSTM and GRU.}
	\vspace{-18pt}
\end{figure}

A significant progress in time series modeling was introduced by  \emph{temporally-aware} ML models, \ie Recurrent Neural Networks (RNNs) (see Figure~\ref{fig:rnns}).
In those, the behaviour of hidden neurons is not only determined by the activations in previous hidden layers, but also by the activations at earlier times.
The activation function for every hidden layer of a RNN is: $h^{(t)} = f(h^{(t-1)}, x^{(t)}, \theta)$.
There, the hidden layer at the time $t$, $h^{(t)}$, is a function of the previous status, $h^{(t-1)}$, of the current input $x^{(t)}$ and of the activation function adopted, $\theta$.
    
The training process of RNNs is usually complex, due to the \emph{unstable gradient problem}: the gradient of the adopted cost function tends to get smaller or bigger as it is propagated back through layers, resulting in a final vanishing or exploding effect, respectively. 
RNNs are unable to model \emph{long term dependencies}, lacking predictive ability when dealing with long sequences of data.
To solve this problem, more eﬀective sequence models are adopted in practical applications, such as Long Short-Term Memory (LSTM) and networks based on the Gated Recurrent Unit (GRU). 
Such \emph{gated RNN} architectures allow the network to accumulate information over a long time period, learning to decide how to forget the old states once that information has been used and processed~\cite{Goodfellow-et-al-2016}.
%Figure~\ref{fig:rnns} depicts these alternatives.

%!TEX root = paper.tex
\vspace{-6pt}
\section{Preliminary Evaluation}\label{sec:eval}
\vspace{-6pt}
We describe here our experimental evaluation of the correlations between cryptocoins.
First we describe our dataset, and then we show several correlation patterns.

\begin{table*}[!t]
	\centering
	\scriptsize
	\begin{center}
		\rowcolors{1}{gray!10}{gray!0}
		\begin{tabularx}{\columnwidth}{lrrrrrr}
			\toprule
			 \rowcolor{gray!25}
			\textbf{Coin} & \textbf{Open} & \textbf{High} & \textbf{Low} &      \textbf{Close} & \textbf{Volume} & \textbf{Market Cap} \\
			\midrule
		 \rowcolor{gray!1}
			BTC  &   29.39/19.55 & 30.18/20.07 & 28.50/18.92 & 29.42/19.53 & 39.63/20.41 & 550.41/368.54 \\
			ETH   & 1.57/1.44 &  1.63/1.49 &  1.51/1.39 & 1.57/1.44 & 20.48/11.05 & 184.19/170.90 \\
			BNB   &  0.20/0.21 & 0.21/0.22 & 0.19/0.20 &  0.20/0.21 & 1.63/1.91 & 33.18/35.39 \\
			XRP   &  5.64e-4/3.94e-4 &  5.90e-4/4.17e-4 & 5.36e-4/3.68e-4 & 5.64e-4/3.94e-4 & 4.59/4.91 & 25.70/17.96 \\
			ADA   & 7.97e-4/8.21e-4 &  8.33e-4/8.55e-4 & 7.60e-4/7.84e-4 & 7.99e-4/8.21e-4 & 2.29/2.77 & 25.67/26.77 \\
			\bottomrule
		\end{tabularx}
	\end{center}
\caption{\label{tab:data-descr}Mean/standard deviation for the top-5 cryptocoins since January 2020 (Open, High, Low and Close expressed in 1K US dollars, Volume and Market Cap in 1B US dollars).\vspace{-12pt}}
\end{table*}

\textbf{Dataset.}\label{subsec:dat}
We collected our dataset from CoinMarketCap~\cite{coinmarketcap}, a leading aggregator of cryptocurrency market data. 
It contains records (High, Low, Open, Close, Volume and Market Capitalization) for 68 coins registered during a time frame of 25 months, namely from 24.12.2019 to 24.01.2022. 
%Those variables are commonly analyzed in order to study the trends of financial instruments over time: more specifically, 
"High" and "Low" are the highest and lowest prices reached by the asset during the considered time frame; "Open" and "Close" the opening and closing market prices; "Volume" the measure of how much it was traded in the last period.
Finally, "Market Capitalization" indicates the total market value of its circulating supply. % \cite{coinmarketcap}.
The dataset includes a total of 51884 observations.  
The resulting time series for each coin trend includes 763 steps.
Table~\ref{tab:data-descr} reports mean and standard deviation for the gathered records and across the top-5 cryptocoins in our dataset.

\begin{figure}[!t]
	\begin{center}
		\includegraphics[scale=0.55]{{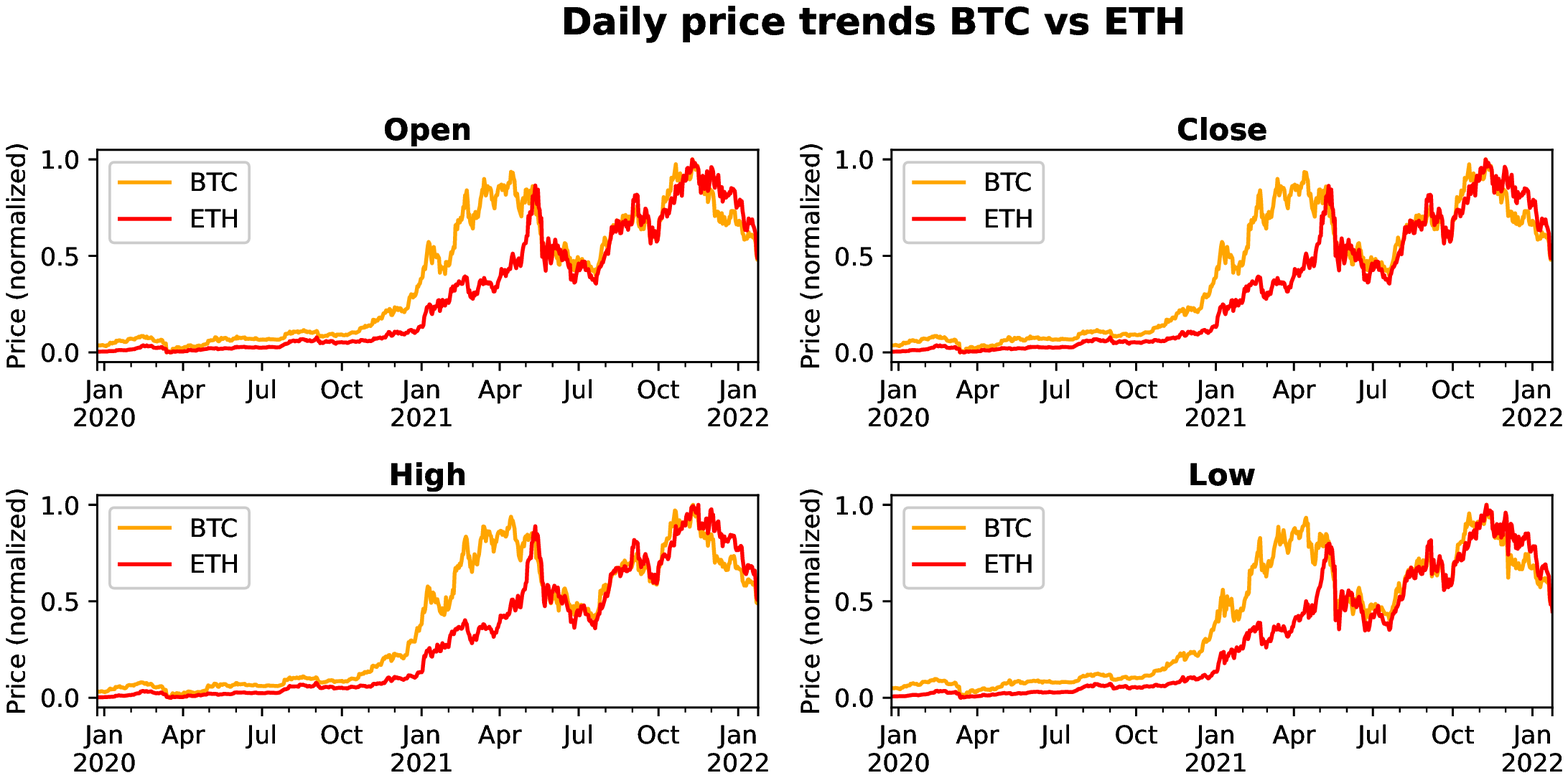}}
	\end{center}
	\vspace{-6pt}
\caption{\label{fig:btc-vs-eth}Trend of Bitcoin and Ethereum prices during the last 2 years.}
	\vspace{-20pt}
\end{figure}

The two major coins in terms of Volume and Market Capitalization are Bitcoin (BTC) and Ethereum (ETH), that we selected as the benchmarks for our subsequent study. 
The price trend of those cryptocoins over the past two years (shown in Figure~\ref{fig:btc-vs-eth}) showed on average a high positive correlation (with a Pearson coefficient $\approx 0.9$). 
%The table below shows in details the average correlation between BTC and ETH computed on all the analyzed variables, including Volume and Market Cap.

%\begin{table}[!h]
%	\begin{center}
%		\begin{tabular}{lr}
%			\toprule
%			Variable & Correlation \\
%			\midrule
%			Open       &  0.9034 \\
%			High       &  0.9027 \\
%			Low        &  0.9048 \\
%			Close      &  0.9031 \\
%			Volume     &  0.7116 \\
%			Market Cap &  0.9029 \\
%			\bottomrule
%		\end{tabular}
%	\end{center}
%\caption{\label{tab:btc-vs-eth}Correlation (Pearson) between BTC and ETH.}
%\end{table}

\begin{figure}[!t]
	\begin{center}
		\includegraphics[scale=1.0]{{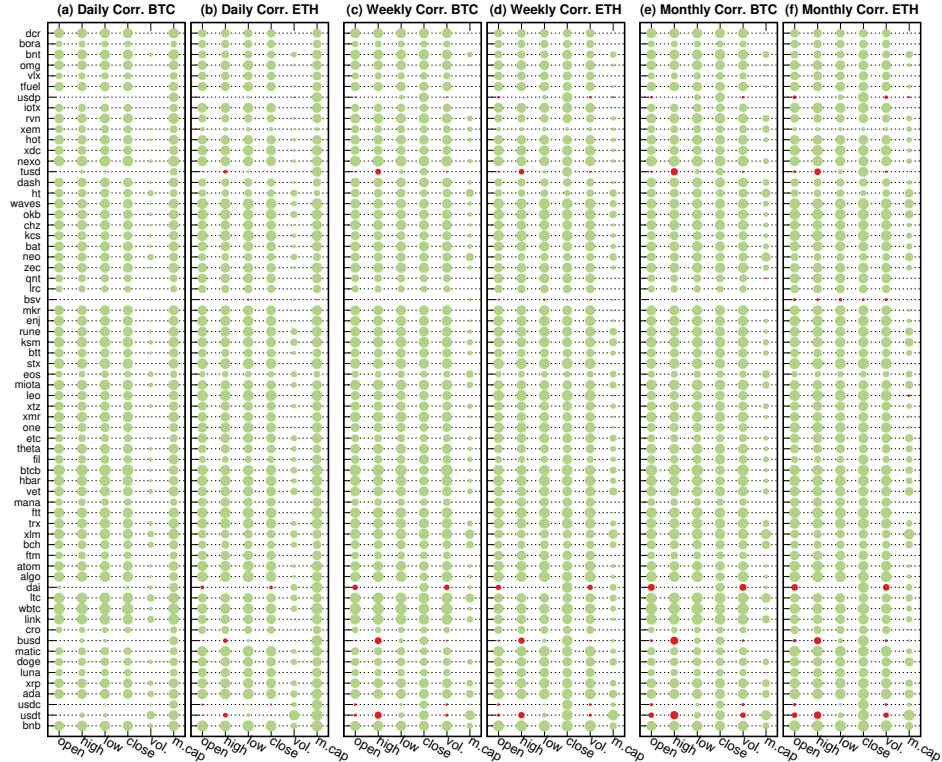}}
	\end{center}
	\vspace{-4pt}
\caption{\label{fig:corr:patterns}Daily, weekly and monthly cross-correlations between the alt-coins and BTC/ETH.}
\vspace{-4pt}
\end{figure}

\textbf{Correlation Patterns.}\label{subsec:corrpatterns}
The aim of the present study is to identify and analyze the presence of cross-correlation patterns in cryptocurrency trends. 
To do so, we analyze the correlations of 66 alt-coins present in our dataset against BTC and ETH, and for three different time frames: daily, weekly and monthly.
For weekly and monthly correlations we define the sequence segments adopting a sliding window approach, where observations are grouped within a window that slides across the data stream. 
The daily observations for each coin are averaged over sliding partitions of 7 and 30 days respectively, and then the correlations with other coins are computed on the resulting aggregated values. 
Note that we postpone the study of \emph{thumbing} windows, where there is no overlapping of data clusters, to future work.
We represent those correlations, averaged among all the studied variables (\ie, High, Low, Open, Close, Volume and Market Cap), as a series of "cross-correlograms" of coins (Figure~\ref{fig:corr:patterns}). 
The radius of each circle represents the strength of the relation (in terms of Pearson coefficient) between each of the considered alt-coin and BTC (Figs.~\ref{fig:corr:patterns}a/c/e) or ETH (Figs.~\ref{fig:corr:patterns}b/d/f).
The color identifies the sign of the correlation (green if positive, red otherwise).
The analysis of the cross-correlogram clearly shows how the vast majority of considered alt-coins are strongly correlated with and follows the same trend of the two market leaders.
Their average values of the Pearson coefficient very close to $1$.
Not surprisingly, the only visible exceptions are represented by the stablecoins available in our dataset (\ie, USDP, TUSD, DAI, BUSD, USDC, USDT), that are pegged to the US dollar and follow standalone trends with total independence from the rest of the coins in the market.
%!TEX root = paper.tex
\vspace{-8pt}
\section{Related Work}\label{sec:relwork}
\vspace{-8pt}
There exists studies which analyzed co-movement and cross-correlation phenomena in cryptocurrency market trends. 
Similar to our study, Katsiampa~\cite{KATSIAMPA2019221} investigated the volatility dynamics of the two major cryptocurrencies, Bitcoin and Ethereum, finding evidence of interdependencies between the two and price responsiveness to major news in the market.
Aslanidis et al.~\cite{ASLANIDIS2019130} showed that cryptocurrencies exhibit similar mean correlation among them, with an unstable trend over time; in addition, the authors computed coins correlation against more traditional assets, detecting an independent behavior respect to other financial markets.
In~\cite{KUMAR20190712},  Bitcoin is identified as the leader in the cryptocurrency market using wavelet-based methods, showing how other coins trends are dependent from BTC price movements: as a result, Bitcoin price drops are immediately reflected in other cryptocurrency prices.
Finally, \cite{CHAUDHARI2020101130} studied the collective behaviour for the cryptocurrency market discovering distinct and not time-persistent community structures characterized by cross-correlation.
%!TEX root = paper.tex
\vspace{-6pt}
\section{Conclusion and Future Work}\label{sec:conclusion}
\vspace{-6pt}
Cryptocoins present very volatile trends on public exchanges.
In this work-in-progress paper, we presented our preliminary evaluation of the correlations between BTC, Ether and 66 other alt-coins.
Our analysis shows strong correlations, suggesting alt-coins follow closely the trends of the two main ones. 
Following this initial study, we will further investigate the cross-correlation between the two market leaders and the alt-coins, in the perspective to forecast their price trends by using the time-series techniques from \S\ref{subsec:tsa}.
We believe that our work could represent a significant starting point for further analyses in co-movement behaviors within the cryptocoin markets and in modeling and forecasting trends of the asset prices. 

Metadata, analysis data, tools and code for reproducibility are available to the research community at \url{https://github.com/quapsale/cryptoanalytics/}.

%\section*{Data and Code}\label{sec:data}
%
%Metadata, analysis data, tools and code for reproducibility are made publicly to the research community at \url{https://github.com/quapsale/cryptocoins-analytics}.
\newpage
\bibliographystyle{splncs04}
\bibliography{paper}

\end{document}